\documentclass[3p,times,twocolumn]{elsarticle}

\usepackage{ecrc}

\volume{00}

\firstpage{1}

\journalname{Nuclear Physics B Proceedings Supplement}

\runauth{A. Pich}

\jid{nuphbp}

\jnltitlelogo{Nuclear Physics B Proceedings Supplement}

\usepackage{amssymb}

 \biboptions{compress}

\usepackage[figuresright]{rotating}

\newcommand{\be}{\begin{equation}}
\newcommand{\ee}{\end{equation}}
\newcommand{\no}{\nonumber}
\newcommand{\bel}[1]{\be\label{#1}}
\newcommand{\ba}{\begin{array}{c}}
\newcommand{\bat}{\begin{array}{cc}}
\newcommand{\ea}{\end{array}}
\newcommand{\beqn}{\begin{eqnarray}}
\newcommand{\eeqn}{\end{eqnarray}}

\newcommand{\bi}{\begin{itemize}}
\newcommand{\ei}{\end{itemize}}

\newcommand{\Br}{\mathrm{Br}}

\begin{document}

\begin{frontmatter}



\dochead{}

\title{TAU2012\ Summary}


\author{Antonio Pich}

\address{Departament de F\'\i sica Te\`orica, IFIC, Universitat de Val\`encia -- CSIC, Apt. Correus 22085, 46071 Val\`encia, Spain}

\begin{abstract}
The main highlights discussed at TAU2012 are briefly summarized. Besides the standard topics on lepton physics covered also at previous conferences (universality, QCD tests, $V_{us}$ determination from $\tau$ decay, $g-2$,
$\nu$ oscillations, lepton-flavour violation), the $\tau$ lepton is playing now a very important role in searches for new physics phenomena.
\end{abstract}

%
%

\end{frontmatter}

\section{Leptonic decays}
\label{sec:lepton}

In the Standard Model (SM) all lepton doublets have identical couplings
to the $W$ boson. Comparing the measured decay widths of leptonic or
semileptonic decays which only differ in the lepton flavour, one can
test experimentally that the $W$ interaction is indeed the same,
i.e. that \ $g_e = g_\mu = g_\tau \equiv g\, $. As shown in
Table~\ref{tab:ccuniv}, the present data verify the universality of
the leptonic charged-current couplings at the 0.2\% level.

The $\tau$ leptonic branching fractions and the $\tau$ lifetime are
known with a precision of $0.3\%$ \cite{Beringer:1900zz}, far away from the
impressive $10^{-6}$ accuracy recently achieved for the muon lifetime
\cite{Webber:2010zf}. The preliminary Belle measurement
$\tau_\tau = 290.18\pm 0.54\pm 0.33$ \cite{Sokolov}
shows that improvements are underway.
The universality tests require also a good determination of the $\tau$ mass,
which is only known at the $10^{-4}$ level.
Uncertainties comparable to the $m_\tau$ world average
are being reached by BES-III, which aims to an accuracy
better than 0.1 MeV \cite{Mo}.

Table~\ref{tab:ccuniv} shows also the contraints obtained from
pion and kaon decays \cite{Pich:2012vd}. The accuracy achieved with
$K_{l2}$ and $K_{l3}$ data is
already competitive with the one obtained from $\tau$ or $\pi_{l2}$ decays.

Owing to the limited statistics available, the leptonic decays of the $W$ boson
only test universality at the 1\% level. At present,
$\Br(W\to\nu_\tau\tau)$ is $2.1\,\sigma/2.7\,\sigma$ larger than
$\Br(W\to \nu_e e / \nu_\mu\mu)$ \cite{Beringer:1900zz}.
This discrepancy cannot be easily explained,
given the stringent limits on $|g_\tau/g_{e,\mu}|$ from $W$-mediated decays \cite{Filipuzzi:2012mg}.

\begin{table}[t] 
\centering \caption{Present constraints on $|g_l/g_{l'}|$.}
\label{tab:ccuniv}
\vspace{0.2cm}
\renewcommand{\tabcolsep}{1.1pc} 
\renewcommand{\arraystretch}{1.15} 
\begin{tabular}{lc}
\hline & $|g_\mu/g_e|$ \\ \hline
$B_{\tau\to\mu}/B_{\tau\to e}$ & $1.0018\pm 0.0014$ \\
$B_{\pi\to \mu}/B_{\pi\to e}$ &  $1.0021\pm 0.0016$ \\
$B_{K\to \mu}/B_{K\to e}$ & $0.9978\pm 0.0020$ \\
$B_{K\to \pi\mu}/B_{K\to\pi e}$ & $1.0010\pm 0.0025$ \\
$B_{W\to\mu}/B_{W\to e}$  & $0.991\pm 0.009$ \\
\hline\hline & $|g_\tau/g_\mu|$  \\ \hline
$B_{\tau\to e}\,\tau_\mu/\tau_\tau$ & $1.0006\pm 0.0021$ \\
$\Gamma_{\tau\to\pi}/\Gamma_{\pi\to\mu}$ &  $0.9956\pm 0.0031$ \\
$\Gamma_{\tau\to K}/\Gamma_{K\to\mu}$ & $0.9852\pm 0.0072$ \\
$B_{W\to\tau}/B_{W\to\mu}$  & $1.032\pm 0.012$
\\ \hline\hline
& $|g_\tau/g_e|$  \\ \hline
$B_{\tau\to\mu}\,\tau_\mu/\tau_\tau$ & $1.0024\pm 0.0021$ \\
$B_{W\to\tau}/B_{W\to e}$  & $1.023\pm 0.011$
\\ \hline
\end{tabular}
\end{table}
%

\section{Hadronic decays}
\label{sec:Hadron}

The $\tau$ is the only known lepton massive enough to decay into hadrons. Its semileptonic decays are ideally suited to investigate the hadronic weak currents and perform
low-energy tests of the strong interaction.

\subsection{The inclusive hadronic width}

The inclusive character of the total $\tau$ hadronic width renders
possible \cite{BNP:92} an accurate calculation of the ratio
$R_\tau \equiv \Gamma [\tau^- \to \nu_\tau \,\mathrm{hadrons}] /
\Gamma [\tau^- \to \nu_\tau e^- {\bar \nu}_e]$.
Its Cabibbo-allowed component can be written as \cite{Pich:2011bb}
\begin{equation}\label{eq:Rv+a}
 R_{\tau,V+A} \, =\, N_C\, |V_{ud}|^2\, S_{\mathrm{EW}} \left\{ 1 +
 \delta_{\mathrm{P}} + \delta_{\mathrm{NP}} \right\}  ,
\end{equation}
where $N_C=3$ is the number of quark colours and
$S_{\mathrm{EW}}=1.0201\pm 0.0003$ contains the
electroweak radiative corrections. 
The non-perturbative contributions are suppressed
by six powers of the $\tau$ mass \cite{BNP:92}
and can be extracted from the invariant-mass distribution of the final
hadrons \cite{LDP:92a}.
The presently most complete and precise experimental analysis, performed with
the ALEPH data, obtains $\delta_{\mathrm{NP}} = -0.0059\pm 0.0014$ \cite{DHZ:05}.

The perturbative QCD contribution $\delta_{\mathrm{P}}$ ($\sim 20\%$) \cite{BNP:92,LDP:92a} is known to $O(\alpha_s^4)$ \cite{BChK:08} and
is very sensitive to $\alpha_s$, allowing for an accurate
determination of the strong coupling \cite{BNP:92}.
The main uncertainty originates in the treatment of higher-order corrections.
The improved renormalization-group resummation of running effects \cite{LDP:92a}
results in smaller values for $\delta_{\mathrm{P}}$ than the naive perturbative truncation of the series at $O(\alpha_s^4)$ \cite{BNP:92,BChK:08}.
While it is well understood than this truncation misses known large corrections \cite{LDP:92a}, giving a poor approximation to $\delta_{\mathrm{P}}$,
it could approach faster the asymptotic renormalon-induced result if the
known $n\le 4$ terms are assumed to be already governed by low-lying infrared renormalons \cite{BJ:08,CF:10}.
Taking the difference between both approaches into account,
the present experimental value $R_{\tau,V+A} = 3.4671\pm 0.0084$ \cite{HFAG} implies \cite{Pich:2011bb}
\begin{equation}\label{eq:alpha-result}
\alpha_s(m_\tau^2)\; =\; 0.329 \pm 0.013\, ,
\end{equation}
significantly larger ($16\,\sigma$) than the result obtained
from the $Z$ hadronic width, $\alpha_s(M_Z^2) =
0.1197\pm 0.0028$ \cite{Beringer:1900zz}.
After evolution up to the scale $M_Z$, the strong
coupling in (\ref{eq:alpha-result}) decreases to
$\alpha_s(M_Z^2)  =  0.1198\pm 0.0015$,
in excellent agreement with the direct measurement at the $Z$ peak.
These two determinations
provide a beautiful test of the predicted QCD running;
i.e., a very significant experimental verification of {\it asymptotic freedom}.

At the presently achieved accuracy, a better experimental assessment of $\delta_{\mathrm{NP}}$ would be welcome,
which requires a more precise determination of the inclusive hadronic distribution. 
This would also allow for an investigation of duality violations \cite{Roig},
which are tiny in $R_\tau$ but more relevant for other moments of the hadronic 
distribution. A recent fit to rescaled OPAL data, 
with moments chosen to maximize duality violations, finds
$\delta_{\mathrm{NP}} = -0.003\pm 0.012$
\cite{Boito:2012cr}, in agreement with the ALEPH result but less precise.

\subsection{$V_{us}$ determination}

The ratio of the $|\Delta S|=1$ and $|\Delta S|=0$ tau decay widths,
$R_{\tau,S}/R_{\tau,V+A}$, provides a very clean determination of $V_{us}$
\cite{Gamiz:2004ar}. To a first approximation, the experimental ratio directly measures $|V_{us}/V_{ud}|^2$.
The small SU(3)-breaking correction $\delta R_\tau$, induced by the strange quark mass,
can be theoretically estimated through a careful QCD analysis \cite{Gamiz:2004ar,Pich:1999hc,Maltman}.
Taking the conservative value $\delta R_\tau = 0.239 \pm 0.030$ \cite{Gamiz},
$R_{\tau,S}=0.1612\pm 0.0028$ \cite{HFAG}
and $|V_{ud}| = 0.97425\pm 0.00022$ \cite{Beringer:1900zz}, one obtains
\beqn\label{eq:Vus_det}
 |V_{us}| &=& \left(\frac{R_{\tau,S}}{\frac{R_{\tau,V+A}}{|V_{ud}|^2}-\delta
 R_{\tau,\mathrm{th}}}\right)^{1/2}
 \no\\ &=&
  0.2173\pm 0.0020_{\mathrm{\, exp}}\pm 0.0010_{\mathrm{\, th}}\, .
\eeqn
This result is lower than the most recent
determination from $K_{l3}$ decays, $|V_{us}|= 0.2238\pm 0.0011$ \cite{Bazavov:2012cd,Cirigliano:2011ny}.
The branching ratios measured by BaBar and Belle are smaller than previous
world averages, which translates into smaller results for $R_{\tau,S}$ and $|V_{us}|$.
Slightly larger central values,
$R_{\tau,S}=0.1653$ and $|V_{us}|=0.2201$,  are obtained
using the $\tau\to\nu_\tau K (\pi)$ branching ratios estimated from
$K\to (\pi)\,\mu\bar\nu_\mu$ \cite{Passemar},
or combining the measured Cabibbo-suppressed $\tau$ distribution with electroproduction data \cite{Maltman}.
Contrary to $K_{l3}$, the final error of the $V_{us}$ determination from
$\tau$ decay is dominated by the experimental uncertainties and, therefore, sizeable
improvements can be expected. Progress on the theoretical side requires a better understanding of the perturbative QCD corrections included in $\delta R_\tau$.

$|V_{us}|$ can also be obtained from exclusive modes, either from the ratio
$\Gamma(\tau\to\nu_\tau K)/\Gamma(\tau\to\nu_\tau \pi)$ or from $\Gamma(\tau\to\nu_\tau K\pi)$, using the appropriate hadronic inputs from lattice calculations ($f_K/f_\pi$, $f_+(0)$). This gives values closer to the $K_{l3}$ result, but with larger errors \cite{HFAG}.

\subsection{Exclusive decays}

A big effort is underway to fully understand the rich pattern of hadronic decay modes of the $\tau$ \cite{Beringer:1900zz,HFAG,Banerjee,Lusiani}. The huge data samples accumulated at the B factories allow for a sizeable reduction of the statistical errors, so systematic uncertainties dominate in most cases.
The decrease of many experimental branching ratios is worrisome.
As pointed out by the PDG \cite{Beringer:1900zz}, 18 of the 20 branching fractions measured at the B factories  are smaller than the previous non-B-factory values. The average normalized difference between the two sets of measurements is $-1.30\,\sigma$. Moreover, the BaBar and Belle results differ significantly for the 6 decay modes measured by both experiments. New measurements and refined analyses are clearly needed.

Recent progress includes the measurement of many high-multiplicity 3- and 5-prong decays
\cite{Lees:2012ks}, modes with $K_S$ [$\pi^-K_S (\pi^0),\pi^-K_S K_S (\pi^0), K^-K_S (\pi^0)$] \cite{Lees:2012de,Sobie,Ryu} and analyses of hadronic distributions
\cite{Ryu,Nugent}. Refined theoretical studies allow for a better understanding of hadronic form factors \cite{Passemar,Dumm}, including the second-class-current decay  $\tau\to\eta\pi\nu_\tau$ \cite{Moussallam}. Forthcoming B-factory analyses and LHC searches will benefit from improved Monte Carlo tools \cite{Ilten} and the incorporation into the TAUOLA library \cite{Shekhovtsova} of the Resonance Chiral Theory constraints \cite{RCHT}.

\section{Anomalous magnetic moments}
\label{sec:g-2}

The most stringent QED test comes from the high-precision
measurements of the $e$ \cite{Hanneke:2008tm} and $\mu$ \cite{Bennett:2006fi}
anomalous magnetic moments
 \ $a_l\equiv (g^\gamma_l-2)/2\, $:
%
\beqn\label{eq:a_e}
 a_e & =& (1\; 159\; 652\; 180.73\pm 0.28) \,\cdot\, 10^{-12}\, ,
 \no\\ \label{eq:a_mu}
 a_\mu &=& (11\; 659\; 208.9\pm 6.3) \,\cdot\, 10^{-10}\, .
\eeqn
The $O(\alpha^5)$ calculation has been completed in both cases
\cite{Aoyama:2012wj}, with an impressive agreement with the measured $a_e$ value.
The dominant QED uncertainty is the input value of $\alpha$, therefore
$a_e$ provides the most accurate determination of the fine structure constant (0.25 ppb),
\be
\alpha^{-1} = 137.035\; 999\; 174 \,\pm\, 0.000\; 000\; 035\, ,
\ee
%
in agreement with the recent (0.66 ppb) measurement from the
atomic $h/m_{\mathrm{Rb}}$ ratio in
${}^{87}$Rb \cite{Bouchendira:2010es}.
%
The heavier muon mass makes $a_\mu$ sensitive to electroweak
corrections from virtual heavier states,
$\delta a_\mu^{\mathrm{ew}} = 15.4\, (0.2)\cdot 10^{-10}$  \cite{Miller:2007kk},
and QCD effects which are at present the main source of uncertainty
\cite{a_mu_had}. There is still a significant difference between the
hadronic vacuum polarization (hvp) corrections extracted \cite{Davier:2010nc} from
$e^+e^-$, $\delta a_\mu^{\mathrm{hvp},e^+e^-} = 692.4\, (4.1)\cdot 10^{-10}$,
and $\tau$ data, $\delta a_\mu^{\mathrm{hvp},\tau} = 701.5\, (4.7)\cdot 10^{-10}$, and discrepancies among different $e^+e^-$ experiments
remain after the most recent BaBar, Belle, CMD-3, KLOE and SND analyses \cite{sig_had,Shwartz}.
Including the so-called light-by-light contributions,
$\delta a_\mu^{\mathrm{lbl}} = 10.5\, (2.6)\cdot 10^{-10}$ \cite{Prades:2009tw},
and NLO hvp corrections, $\delta a_\mu^{\mathrm{hvp,NLO}} = -9.8\, (0.1)\cdot 10^{-10}$ \cite{Hagiwara:2011af}, the final SM prediction
\bel{eq:QED_pred}
a_\mu^{\mathrm{th}}  = \left\{ \bat
(11\, 659\, 180.4\pm 4.9)\cdot 10^{-10}
&\; (e^+e^-) \\
(11\, 659\, 189.5\pm 5.4)\cdot 10^{-10}
&\; (\tau)
\ea \right.
\ee
%
differs from the experimental value by $3.6\,\sigma$
($e^+e^-$) \ or \ $2.3\,\sigma$ \ ($\tau$). New precise $e^+e^-$ and
$\tau$ data sets are needed to settle the true value of
$a_\mu^{\mathrm{th}}$.
Improved predictions are needed to match the aimed $10^{-10}$ accuracy of the
proposed muon experiments at Fermilab and J-PARC \cite{Lee}.

With a predicted value $a_\tau^{\mathrm{th}}= 117\, 721\, (5)\cdot 10^{-8}$ \cite{Eidelman:2007sb}, the $\tau$ anomalous magnetic moment has an enhanced sensitivity to new physics because of the large tau mass. However, it is essentially unknown experimentally:
$a_\tau^{\mathrm{exp}}= -0.018\pm 0.017$ \cite{Abdallah:2003xd}.

\section{CP violation}
\label{sec:CPV}

A variety of CP-violating observables (rate, angular and polarization asymmetries, triple products, Dalitz distributions, etc.) can be exploited to search for violations of the CP symmetry in $\tau$ decay and/or production \cite{Bigi}. While the SM predictions are very small, new-physics signals could show up in the $\tau$ data.

The $\tau^+\to\pi^+K_S\bar\nu_\tau\, (\ge 0 \pi^0)$ rate asymmetry recently measured by BaBar \cite{Sobie,BABAR:2011aa},
\be
\mathcal{A}_\tau\equiv\frac{\Gamma-\bar\Gamma}{\Gamma+\bar\Gamma} =
(-0.36\pm 0.23\pm 0.11)\%\, ,
\ee
differs by $2.8\,\sigma$ from the expected value due to $K^0$--$\bar K^0$ mixing, $\mathcal{A}_\tau = (0.36\pm 0.01)\%\,$ \cite{Bigi:2005ts,Grossman:2011zk}.
Belle has also searched for a CP signal in this decay mode through a difference in the $\tau^\pm$ angular distributions, finding a null result
at the 0.2--0.3\% level \cite{Bischofberger:2011pw}.

\section{Tau production in B decays}
\label{sec:B}

An excess of events in two $b\to c\,\tau^-\bar\nu_\tau$ transitions has been reported by BaBar \cite{Lees:2012xj}.
Including the previous Belle measurements~\cite{Bozek:2010xy} ($\ell=e,\mu$),
\beqn\label{eq:Babar}
 R(D)&\!\! \equiv &\!\!
\frac{\Br(\bar B\to D\tau^-\bar\nu_\tau)}{\Br(\bar B\to D\ell^-\bar\nu_\ell)}
\, = \, 0.438\pm0.056\, ,
\nonumber\\  \\  
R(D^*)&\!\! \equiv &\!\!
\frac{\Br(\bar B\to D^*\tau^-\bar\nu_\tau)}{\Br(\bar B\to D^*\ell^-\bar\nu_\ell)}
\, = \, 0.354\pm0.026\, .
\nonumber\eeqn
The SM expectations, $R(D)=0.296\pm 0.016$ and $R(D^*)=0.252\pm 0.003$~\cite{Fajfer:2012jt,Sakaki:2012ft}, are significantly lower. If confirmed, this could signal new-physics contributions violating lepton-flavour universality.

A sizable deviation from the SM was previously observed in 
$B^-\to\tau^-\bar\nu_\tau$. However, Belle~\cite{Adachi:2012he} finds now
a much lower value in agreement with the SM; combined with the BaBar result~\cite{Lees:2012ju}, gives the average $\Br(B^-\to\tau^-\bar\nu_\tau) = (1.15\pm 0.23)\cdot 10^{-4}$, to be compared with the SM expectation
$(0.733\,{}^{+\,0.121}_{-\,0.073})\cdot 10^{-4}$~\cite{Charles:2004jd}.

These results are intriguing enough to trigger the theoretical interest.
The enhancement of  $\tau$ production could be generated by new physics contributions with couplings  proportional to fermion masses. In particular, it could be associated with the exchange of a charged scalar within two-Higgs-doublet models. Although the Babar data rules out the usually adopted ``Type II'' scenario~\cite{Lees:2012xj,Fajfer:2012jt}, these measurements can be accommodated \cite{Celis:2012dk} by the more general framework of the ``Aligned Two-Higgs-Doublet Model''~(A2HDM)~\cite{Pich:2009sp}, albeit creating a tension with charm data.

\section{Lepton-flavour violation}
\label{sec:LFV}

We have clear experimental evidence that neutrinos are
massive particles and there is mixing in the lepton sector.
The solar, atmospheric, accelerator and reactor neutrino
data, lead to a consistent pattern of oscillation parameters
\cite{Beringer:1900zz}. The main recent advance is the establishment of
a sizeable non-zero value of $\theta_{13}$, both in accelerator (T2K, Minos)
and reactor experiments (Double-Chooz, Daya Bay, Reno) \cite{theta13},
with a statistical significance which reaches the $7.7\,\sigma$ at Daya Bay~\cite{An:2012bu}:
\be
\sin^2{2\theta_{13}}\, =\, 0.089\pm 0.010\pm 0.005\, .
\ee
This increases
the interest for a next-generation of long-baseline $\nu$  
experiments to measure the CP-violating phase $\delta$ and resolve the neutrino mass hierarchy \cite{Rubbia}.

Other neutrino highlights presented at this conference include the
second $\nu_\mu\to\nu_\tau$ candidate reported by OPERA \cite{opera},
and the IceCube search for ultra-high energy $\nu_\tau$, finding
three events which are statistically consistent with background
fluctuations \cite{IceCube}.

The smallness of neutrino masses implies a strong suppression of
lepton-flavour violation (LFV) in charged lepton decays, which can be
avoided in models with sources of LFV not related to $m_{\nu_i}$.
LFV processes have the potential to probe physics at scales much higher than the TeV. The LFV scale can be
constrained imposing the requirement of a viable leptogenesis.
Recent studies within different new-physics scenarios find interesting
correlations between $\mu$ and $\tau$ LFV decays,
with $\mu\to e\gamma$ often expected to be close to the present exclusion limit
\cite{Hisano}.

The B Factories are pushing the experimental limits on neutrinoless LFV
$\tau$ decays to the $10^{-8}$ level \cite{Hayasaka},
increasing in a drastic way the sensitivity to new physics scales.
A rather competitive upper bound on $\tau\to 3\mu$ has been also obtained
at LHCb \cite{Harrison}.
Future experiments could improve further some limits to the $10^{-9}$
level \cite{Inami}, allowing to explore interesting and  totally unknown phenomena.

Complementary information is provided by the MEG experiment, which
has already set a limit on $\mu^+\to e^+\gamma$ five times tighter
than previous experiments and aims to reach
a sensitivity of $10^{-13}$ \cite{MEG,Natori}.
A possible $10^{4}$ improvement in $\mu\to 3 e$, reaching a sensitivity of $10^{-16}$, is also under study at PSI \cite{Berstein},
and ongoing projects at J-PARC \cite{Nishiguchi} and FNAL
\cite{Berstein} aim to study $\mu\to e$
conversions in muonic atoms, at the $10^{-16}$ level.

Lepton-number violation has also been tested in $\tau$
($\tau^-\to (e/\mu)^+ h^- h^-, \Lambda\pi^-, \bar p \gamma$ \cite{Beringer:1900zz},
$\tau^-\to p \mu^-\mu^-, \bar p \mu^+\mu^-$ \cite{Harrison})
and meson ($M\to h l^- l'^-$ \cite{Santos}) decays with sensitivities approaching in some cases the $10^{-8}$ level. These bounds
constrain models of new physics
involving Majorana neutrinos with masses in the GeV range \cite{Lopez-Castro}.

\section{Tau physics at the LHC}
\label{sec:LHC}

Owing to their high momenta, tightly collimated decay products and low multiplicity, $\tau$ leptons provide excellent signatures to probe new physics at high-energy colliders \cite{signatures}.  Moreover,
the distribution of the $\tau$ decay products contains precious polarization information.
The $\tau$ signal has been already exploited successfully at the LHC to measure $W$, $Z$ and top production cross sections ($W^-\to\tau^-\bar\nu_\tau$, $Z\to\tau^+\tau^-$, $t\to b\tau^+\nu_\tau$) \cite{Ilten,Ferro}, and ATLAS has reported the first $\tau$ polarization measurement ever made at hadron colliders, using the $\tau\to 2\pi\nu_\tau$ decay in
$W^-\to\tau^-\bar\nu_\tau$ \cite{:2012cu}.

The $\tau$ is the heaviest lepton coupling to the Higgs; with $m_H = 126$~GeV, the decay $H\to\tau^+\tau^-$ has the fourth largest Higgs branching ratio.
No significant $H\to\tau^+\tau^-$ signal has been found up to now.
The experimental analyses are quantified in terms of the signal-strength parameter, measuring the product of Higgs production cross section and branching ratio, normalized to the SM prediction. In the $H\to\tau^+\tau^-$ mode ATLAS quotes $\mu_{\tau\tau} = 0.8\pm 0.7$ \cite{Higgs_ATLAS}, while CMS finds $\mu_{\tau\tau} = 0.7\pm 0.5$ \cite{Higgs_CMS}.
These values are consistent with either the SM or the absence of a $H\tau^+\tau^-$ coupling.

Present searches for new phenomena, taking advantage of the $\tau$ signal,
include bounds on $Z'$ bosons ($Z'\to \tau^+\tau^-$),
supersymmetric neutral ($H\to\tau^+\tau^-$) and charged ($t\to H^+b\to \tau^+\nu_\tau b$) Higgses \cite{Benitez},
and the BaBar constraints on a light CP-odd neutral scalar ($\Upsilon\to\gamma A^0\to\gamma\tau^+\tau^-$) \cite{Banerjee,Lees:2012te}. Significant improvements are to be expected with the increasing LHC luminosity and the use of more refined tools, such as charge asymmetries \cite{Valencia}, to disentangle different new-physics scenarios.

\section{Outlook}
\label{sec:outlook}

While the $\tau$ lepton continues being an increasingly precise laboratory to perform relevant tests of QCD and the electroweak theory, this conference has witnessed the opening of a new era with this heavy lepton becoming now a superb tool in searches for new phenomena. The ongoing LHC programme will be complemented with refined low-energy measurements at Belle-II \cite{Inami}, Bes-III \cite{Mo} and, perhaps, a future Super Tau-Charm Factory \cite{Shwartz}, and muon experiments \cite{Lee,Natori,Berstein,Nishiguchi}. There is an exciting future ahead of us and unexpected surprises may arise, probably establishing the existence of new physics beyond the SM and offering clues to the problems of mass generation, fermion mixing and family replication.


\section*{Acknowledgements}
I want to thank the local organizers for this enjoyable conference.
This work has been supported by the Spanish Government
and EU ERDF funds 
[grants FPA2007-60323, FPA2011-23778 and CSD2007-00042]. 







\end{document}